# The proposed use of ion beams or cosmic rays as a new stimulus technique in the search for biological nonlocality

Fred H. Thaheld

fthaheld@directcon.net

**Abstract**

Research over the last several decades appears to suggest the possibility of the existence of biological nonlocality between human subjects. This is based on photostimulation and resulting EEG correlations between stimulated and non-stimulated subjects. Replication of results has proven to be difficult due to the low microvoltages generated and the lack of clarity in the correlated brain waves. It is proposed to use either heavy ion beams from accelerators or cosmic rays as alternate stimulus sources. This is based on existing research derived from irradiating mice and brain tumor patients with heavy ions and with astronauts exposed to cosmic rays.

A *counterintuitive* body of research has been slowly accumulating over the last several decades which *seems* to suggest the possibility of the existence of biological nonlocality between human subjects [1,2]. This can be considered as the biological equivalent, or analogous to, the original physics experiment by Aspect, which revealed the existence of quantum mechanical nonlocality, by passing one photon of an entangled pair through a polarizer, and having the second entangled photon respond in a similar polarized fashion [3]. I.e., a measurement was made on the first photon and conveyed in nonlocal fashion to the second photon which was several meters distant.

The usual research technique involving the humans, has been to subject one of a pair of individuals to regular or patterned photostimulation, thereby altering their electroencephalogram (EEG), and observing correlated EEG effects in the non-stimulated subject who is located several meters away, usually in a separate room or Faraday cage [1,2]. Other related research has involved recording the fMRI of a non-stimulated subject. The problem with these experiments has been the limited number of subjects involved and the lack of replication, due to the low microvoltages generated by the non-stimulated subjects in their EEGs in comparison to the microvoltages in the EEGs of the stimulated subjects. That is one of the reasons that the author has proposed some additional stimulus techniques over and above photostimulation, such as transcranial magnetic stimulation (TMS), where one gets several different and simultaneous EEG responses in addition to the appearance of *phosphenes* [1-2,4]. For the purposes of this proposal it is important to mention here that in a similar vein, it has been noticed that patients undergoing ion therapy for brain tumors, simultaneously observe *phosphenes* [5,6]. Astronauts in space also observe *phosphenes* or *light flashes* (LF) as a result of cosmic rays impinging on the retina [6].

Experiments have also been conducted with neurons adhering to two separate microelectrode arrays (MEAs), both of which are in Faraday cages, where one MEA is stimulated with a laser and correlated electrical signals *appear* in the non-stimulated MEA [7,8]. Future experiments are planned with more stringent shielding to rule out any possible ultraweak electromagnetic influences in the case of neurons grown on MEAs.

In addition to proposing experiments involving the use of human subjects, the author has proposed the stimulation of monkeys and dolphins with patterned photostimulation and TMS [2]. He would now like to broaden the stimulus technique to include ions from accelerators and cosmic rays utilizing both human subjects and mice.

In the case of mice we would have one pair, with both hooked up to separate electroretinogram (ERG), EEG and visual evoked potential (VEP) [9]. Only one of the pair is irradiated with $^{12}C$ ions or is stimulated with light from a white LED, either simultaneously or separately. We now want to determine if the other mouse, which is several meters distant or in an adjacent room, reveals any simultaneous correlated or anomalous responses or overlap in the areas of the ERG, EEG or VEP. For initial experimental purposes there would be no need to have the non-stimulated mouse in a Faraday cage to rule out any possible electromagnetic influences. It is only if we observe some type of correlations that one could proceed to this step. Any evidence for any type of correlation over the period of ion bombardment would be evidence for biological nonlocality.

Essentially the same arrangement could be made for patients undergoing radiation therapy for tumors, using a pair with both hooked up to separate ERGs, EEGs and VEPs and, with a white LED for separate or simultaneous light stimulation. Once again they are separated by several meters without a Faraday cage i.e., the non-stimulated subject can be in an adjacent room. We want to determine if, while one of them is being subjected to ion radiation, we get some correlated response in either the ERG, EEG or VEP of the non-stimulated subject. In addition, we will want to see if the non-stimulated subject *also* reports on seeing any *phosphenes*! If any correlations are observed one can then place the non-stimulated subject in a Faraday cage to rule out any extraneous influences.

The final proposal has to do with astronauts in space and human subjects on the ground. At the present time astronauts on the International Space Station (ISS) are utilizing what is referred to as the Anomalous Long Term Effects on Astronauts (ALTEA) facility (6,9). ALTEA consists of a silicon detector system (SDS) positioned around the astronaut's head on a helmet shaped holder containing a 32 electrode EEG cap, including 3 floating electrodes for ERG measurements. It also has a visual stimulator unit (VSU) which delivers the light stimulation paradigm for the VEP. The VSU can also be used at any time to create additional stimuli in conjunction with that arising from the cosmic rays. The SDS is able to reconstruct the trajectories of specific cosmic ray particles and ascertain what type, while correlating this with electrophysiological readings such as ERG, EEG and VEP. ALTEA measures the particles passing through the astronaut's eyes/brain, their electrophysiological brain dynamics and the visual system status and, each perception of a LF is signaled with a pushbutton.

On any given day thousands of EEGs are being taken around the world. The objective will be to see if we can observe any correlations of an ERG, EEG, VEP or LF nature between the astronauts and subjects undergoing EEGs on earth, just as in the previous mice and tumor patient experiments. Since the ISS is in motion around the earth at an altitude of 350 miles, any ERGs, EEGs, VEPS and LF on the earth would have to be coordinated in real time with the ERGs, EEGs, VEPs and LF originating from the ISS. This should be able to be done no matter where the ISS is in its orbit with regards to any chosen site on earth.

The argument can be made that if we are having difficulty observing any of these effects on earth, where the participants are only a few meters apart, that it will be many times over more difficult to observe these effects between subjects that are several hundred to several thousand miles apart! Except that in the case of entangled photon experiments, nonlocality has been observed where the photons have been over 10 km apart and more [10]. Distance is no criteria where nonlocality is involved as has already been amply demonstrated [3].

We do not know what effect the heavy ion beams on earth or the cosmic rays in space will have as a stimulus but, it could be of an unusual and unexpected nature. It is of interest to note here that the heavy ion beam intensity at GSI/Darmstadt is of the order of $10^{7-8}$ ions $s^{-1}$, and that ALTEA has had a collaboration with GSI/Heidelberg to measure the EEG of the patients while they are being irradiated [6]. One cannot begin to predict what effect this will have on the nervous system. As regards the cosmic rays as a stimulus source, they are high energy charged particles that travel at nearly the speed of light, and are the nuclei of atoms ranging from the lightest to the heaviest in the periodic table. They also include high energy electrons, positrons and other subatomic particles such as muons and pions. Approximately 89% of the nuclei are H (protons), 10% He and 1% heavier elements [11,12].

In addition, the rare possibility of ultra-high energy cosmic rays (UHECRs), with energies of $10^{18}$-$10^{20}$ eV and above, impinging upon the retina or the brain/visual cortex and eliciting an unusual ERG, EEG, VEP or LF, cannot be ruled out [13]. This might be the analogous equivalent of the rare events which are recorded in neutrino, super heavy element and collider experiments where, from trillions of bombarding particles one is lucky to get 2-3 recordable events over a period of days or weeks. We are proceeding into uncharted territory and should be prepared to see events of an electrophysiological nature which, even if they are rare, may dramatically stand out, thereby more than making up for their rarity. And, we should be prepared that certain rare individuals may be capable of revealing this effect.

Lest you think that this proposal might be stretching credulity to the extreme, for several decades now scientists have been conducting a Search for Extraterrestrial Intelligence (SETI) by monitoring and analyzing radio signals emanating from space. This search is now being pursued under the auspices of what is referred to as SETI@home [14]. This is a distributed or grid computing project using internet-connected computers to analyze radio signals to determine if they come from intelligent life outside the earth. At the present time over 250,000 computers around the world are engaged in this search, with over $10^{16}$ bytes (10 petabytes!) having been analyzed so far, with an input from the radio signal receiving dishes of 5 x $10^{6}$ bits/sec. While the viability and practicality of the distributed grid computing concept has been proven, to date no evidence for extraterrestrial intelligence signals has been shown via SETI@home.